\documentclass[10pt]{article}
\usepackage{amsfonts}
\usepackage{amsmath,amssymb}
\usepackage{graphicx}
\usepackage{caption2}
\usepackage{epsfig}
\usepackage{subfigure}
\usepackage[dvips]{color}

\def\dref#1{(\ref{#1})}
\def\rm{\mathrm}
\begin{document}

\begin{center}
{\Large\bf{Containment Control of Linear Multi-Agent Systems
with Multiple Leaders of Bounded Inputs Using Distributed Continuous Controllers}}
\footnote[1] {\small Zhongkui Li and
Zhisheng Duan are with the State Key Laboratory for Turbulence
and Complex Systems,
Department of Mechanics and Engineering Science, College of Engineering,
Peking University, Beijing 100871, China
(E-mail: zhongkli@gmail.com; duanzs@pku.edu.cn).
Wei Ren is with
the Department of Electrical Engineering, University of California, Riverside, CA, 92521, USA
(E-mail: ren@ee.ucr.edu).
Gang Feng is with the Department of Mechanical and Biomedical Engineering,
City University of Hong Kong, China
(Email: megfeng@cityu.edu.hk)}
\end{center}

\vskip 0.6cm
\centerline{Zhongkui Li, Zhisheng Duan, Wei Ren,
Gang Feng}

\vskip 1cm

{\noindent \small {\bf  Abstract}: This paper considers the
containment control problem for
multi-agent systems with general linear dynamics and
multiple leaders whose control inputs
are possibly nonzero and time varying. Based on the relative states
of neighboring agents,
a distributed static continuous
controller
is designed,
under which the containment error is uniformly ultimately
bounded and the upper bound
of the containment error can be made arbitrarily
small, if
the subgraph associated with
the followers is undirected and for each
follower there exists at least one leader that has a directed path
to that follower.
It is noted that
the design of the static controller
requires the knowledge of
the eigenvalues of the Laplacian matrix
and the upper bounds of the leaders'
control inputs.
In order to remove these requirements,
a distributed adaptive continuous controller
is further proposed, which can be
designed and implemented by each follower
in a fully distributed fashion.
Extensions to the case where only local output
information is available are discussed.

\vskip 0.2cm

}

\vskip 0.6cm

\section{Introduction}

Consensus is a fundamental problem
in the area of cooperative control
of multi-agent systems
and has attracted a lot of interest
from the systems and control community
in the last decade.
Consensus
means that a group of agents reaches an agreement on a physical
quantity of interest by interacting with their local neighbors.
For
recent advances
of the consensus
problem, readers are referred to
\cite{olfati-saber2004consensus,ren2007information,abdessameud2010consensus,
li2010consensus,li2011dynamic,
zhang2011optimal,zhang2011fast,you2011network,li2011trackingTAC,li2012adaptiveauto,grip2012output}
and references
therein. Roughly speaking,
existing
consensus algorithms can be categorized into two classes,
namely, consensus without a leader and
consensus with a leader. The case of consensus
with a leader is also called leader-follower consensus or distributed
tracking.

The distributed tracking problem deals with only one leader.
However, in some practical applications, there might exist more than one leader in
agent networks.
In the presence of multiple leaders, the containment control problem
arises, where the followers are to be driven into a given geometric space spanned by the leaders
\cite{ji2008containment}.
The study of containment control has been motivated by many
potential applications. For instance, a group of autonomous
vehicles (designated as leaders) equipped with necessary sensors to detect the hazardous obstacles
can be used to safely maneuver another group of vehicles (designated as followers) from one target
to another, by ensuring that the followers are contained within the moving safety area formed by the
leaders \cite{ji2008containment,cao2011distributed2}. A hybrid containment control law is proposed
in \cite{ji2008containment} to drive the followers into
the convex hull spanned by the leaders. Distributed containment control problems are studied in
\cite{cao2009containment,cao2011distributed2,cao2012distributed} for a group of first-order
and second-order integrator agents under fixed and
switching directed communication topologies.
The containment control is considered
in \cite{lou2012target} for second-order multi-agent systems
with random switching topologies.
A hybrid model predictive control scheme is proposed in
\cite{galbusera2013hybrid}
to solve the containment and distributed
sensing problems
in leader/follower multi-agent systems.
The authors in
\cite{mei2012distributed,dimarogonas2009leader,meng2010distributed}
study the containment
control problem for a collection of Euler-Lagrange
systems. In particular, \cite{dimarogonas2009leader} discusses the case with multiple
stationary leaders, \cite{meng2010distributed} studies the case of dynamic leaders with finite-time convergence,
and \cite{mei2012distributed} considers the case with parametric uncertainties.
In the above-mentioned works,
the agent dynamics are assumed to be single, double
integrators, or second-order Euler-Lagrange systems,
which might be restrictive in some circumstances.
The containment control for multi-agent systems
with general linear dynamics is considered
in \cite{li2011containment}, which however
assumes
the leaders' control inputs to be zero.
In many cases,
the leaders might need
nonzero control actions
to regulate their state trajectories,
e.g., to avoid obstacles or
to form a desirable safety area.

In this paper,
we study the
distributed containment control problem for
multi-agent systems with general linear dynamics and
multiple leaders whose control inputs
are possibly nonzero and time varying.
Based on the relative states of neighboring agents,
a distributed discontinuous controller
is designed to ensure that
the containment error asymptotically converges
to zero, if
the subgraph associated with
the followers is undirected and for each
follower there exists at least one leader that has a directed path
to that follower. It is pointed out that the discontinuous controller may
cause the undesirable chattering phenomenon in real implementation.
To eliminate the chattering effect, using
the boundary layer concept,
a static continuous
containment controller is then constructed,
under which the containment error is uniformly ultimately
bounded and the upper bound
of the containment error can be made arbitrarily
small.
It is noted that
the design of this static containment controller
requires
the knowledge of the eigenvalues of the Laplacian matrix
and the upper bounds of the leaders'
control inputs.
In order to remove these requirements,
a distributed adaptive continuous containment controller
is further proposed.
A distinct feature of the proposed adaptive containment
controller is that it can be
designed and implemented by each follower
in a fully distributed fashion
without requiring any global information.
Extensions to the case where only local output information
is available are discussed.
Based relative estimates of the states of neighboring agents,
distributed observer-based containment controllers are proposed.
A sufficient condition for the existence of these
containment controllers is that each agent is stabilizable and
detectable.

Compared to the previous works
\cite{ji2008containment,cao2009containment,cao2011distributed2,cao2012distributed,lou2012target,
mei2012distributed,dimarogonas2009leader,meng2010distributed,li2011containment},
the contribution of this paper is at least three-fold.
First, in contrast
to \cite{ji2008containment,cao2009containment,cao2011distributed2,cao2012distributed,lou2012target,
mei2012distributed,dimarogonas2009leader,meng2010distributed}
which puts restrictions on the agent dynamics
and \cite{li2011containment} which assumes
the leaders' control inputs to be zero,
the results obtained in this paper
are applicable to multi-agent systems with general linear
dynamics and
multiple leaders whose control inputs
are possibly nonzero and bounded.
Second,
contrary to the discontinuous
controllers in \cite{cao2009containment,cao2011distributed2,cao2012distributed,meng2010distributed,
mei2012distributed},
a distinct feature
of the proposed containment
controllers
is that they are continuous,
for which case the undesirable chattering
phenomenon can be avoided.
It is worth mentioning that
with the discontinuous functions
replaced with the continuous ones,
it is no longer clear how the controllers and the adaptive gain design will function.
It is hence challenging to analyze and show the ultimate boundedness
of the containment errors
and the adaptive coupling gains
using the proposed continuous controllers.
Third, the adaptive containment
controllers proposed in this paper
can be implemented in a fully distributed
fashion without requiring
any global information.

The rest of this paper is organized as follows. Some useful results of the
graph theory and ultimate boundedness
are reviewed in Section 2. The containment control problem
is formulated and discontinuous
containment controllers are proposed
in Section 3. Distributed static and adaptive continuous containment controllers
based on the relative state information
are considered in Section 4.
Extensions to the case with output feedback controllers
are discussed in Section 5.
Simulation
examples are presented in Section 6 to illustrate the analytical results. Conclusions are drawn in
Section 7. 

\section{Mathematical Preliminaries}

\subsection{Notations}
Let $\mathbf{R}^{n\times n}$
be the set of $n\times n$ real matrices.
The superscript $T$ means
transpose for real matrices. $I_N$ represents the identity matrix of
dimension $N$.
Denote by $\mathbf{1}$ a column vector
with all entries equal to one. $\rm{diag}(A_1,\cdots,A_n)$
represents a block-diagonal matrix with matrices $A_i,i=1,\cdots,n,$
on its diagonal. For real symmetric matrices $X$
and $Y$, $X>(\geq)Y$ means that $X-Y$ is positive (semi-)definite.
$A\otimes B$ denotes the Kronecker product of matrices $A$ and $B$.
For a vector $x\in\mathbf{R}^n$, let
$\|x\|$ denote its 2-norm.
For a symmetric matrix $A$, $\lambda_{\min}(A)$ and $\lambda_{\max}(A)$
denote, respectively, the minimum and maximum eigenvalues
of $A$.
A matrix is Hurwitz (stable) if all of its eigenvalues have strictly negative
real parts.

\subsection{Graph Theory}

A directed graph $\mathcal {G}$ is a pair $(\mathcal {V}, \mathcal
{E})$, where $\mathcal {V}=\{v_1,\cdots,v_N\}$ is a nonempty finite
set of nodes and $\mathcal {E}\subseteq\mathcal {V}\times\mathcal
{V}$ is a set of edges, in which an edge is represented by an
ordered pair of distinct nodes. For an edge $(v_i,v_j)$, node $v_i$
is called the parent node, node $v_j$ the child node, and $v_i$ is a
neighbor of $v_j$. A graph with the property that
$(v_i,v_j)\in\mathcal {E}$ implies $(v_j, v_i)\in\mathcal {E}$ for
any $v_i,v_j\in\mathcal {V}$ is said to be undirected. A path from
node $v_{i_1}$ to node $v_{i_l}$ is a sequence of ordered edges of
the form $(v_{i_k}, v_{i_{k+1}})$, $k=1,\cdots,l-1$. A subgraph
$\mathcal {G}_s=(\mathcal {V}_s, \mathcal {E}_s)$ of $\mathcal {G}$
is a graph such that $\mathcal {V}_s\subseteq\mathcal {V}$ and
$\mathcal {E}_s\subseteq \mathcal {E}$.
A directed graph contains a directed spanning tree if
there exists a node called the root, which has no parent node, such
that the node has directed paths to all other nodes in the graph.

The adjacency matrix $\mathcal {A}=[a_{ij}]\in\mathbf{R}^{N\times
N}$ associated with the directed graph $\mathcal {G}$ is defined by
$a_{ii}=0$, $a_{ij}=1$ if $(v_j,v_i)\in\mathcal {E}$ and $a_{ij}=0$
otherwise. The Laplacian matrix $\mathcal {L}=[\mathcal
{L}_{ij}]\in\mathbf{R}^{N\times N}$ is defined as $\mathcal
{L}_{ii}=\sum_{j\neq i}a_{ij}$ and $\mathcal {L}_{ij}=-a_{ij}$,
$i\neq j$. For undirected graphs, both $\mathcal {A}$ and $\mathcal
{L}$ are symmetric. It is easy to see that
zero is
an eigenvalue of $\mathcal {L}$ with $\mathbf{1}$ as a corresponding
right eigenvector and all nonzero eigenvalues have positive real
parts. Furthermore, zero is a simple eigenvalue of $\mathcal {L}$ if
and only if $\mathcal {G}$ has a directed spanning tree
\cite{agaev2005spectra,ren2005consensus}.

\section{Problem Formulation and Discontinuous Containment Controllers}

Consider a group of $N$ agents with general continuous-time
linear dynamics, described by
\begin{equation}\label{1c}
\begin{aligned}
    \dot{x}_i &=Ax_i+Bu_i,\\
    y_i & = Cx_i, \quad i=1,\cdots,N,
\end{aligned}
\end{equation}
where $x_i\in\mathbf{R}^n$, $u_i\in\mathbf{R}^{p}$, and
$y_i\in\mathbf{R}^{q}$ are, respectively, the state, the control
input, and the output of the $i$-th agent, and $A$, $B$,
$C$, are constant matrices with compatible dimensions.

In this paper, we consider the case where there exist
multiple leaders. Suppose that there
are $M$ ($M<N$) followers and $N-M$ leaders. An agent is called a
leader if it has no neighbor
and is called a follower
if it has at least one neighbor. Without loss of generality,
we assume that the agents indexed by $1,\cdots,M$, are followers,
while the agents indexed by $M+1,\cdots,N$, are leaders. We use $\mathcal
{R}\triangleq\{M+1,\cdots,N\}$ and $\mathcal
{F}\triangleq\{1,\cdots,M\}$ to denote, respectively, the leader set
and the follower set.
The communication
graph among the $N$ agents is represented by a directed graph
$\mathcal {G}$, which satisfies the following assumption.

{\bf Assumption 1}. The subgraph $\mathcal {G}_s$ associated with
the $M$ followers is undirected. For each follower, there
exists at least one leader that has a directed path to that
follower.

Denote by $\mathcal {L}$ the Laplacian matrix associated with
$\mathcal {G}$. Because the leaders have no neighbors,
it is easy to see that
$\mathcal {L}$
can be partitioned as
\begin{equation}\label{lapc}
\mathcal {L}=\begin{bmatrix}\mathcal {L}_1 & \mathcal {L}_2\\
0_{(N-M)\times M} & 0_{(N-M)\times (N-M)}\end{bmatrix},
\end{equation}
where $\mathcal {L}_1\in\mathbf{R}^{M\times M}$ is symmetric and
$\mathcal {L}_2\in\mathbf{R}^{M\times (N-M)}$.

{\small \bf Lemma 1} \cite{meng2010distributed}.
Under Assumption 1, all the eigenvalues of
$\mathcal {L}_1$ are positive, each entry of $- \mathcal
{L}_1^{-1}\mathcal {L}_2$ is nonnegative, and each row of
$-\mathcal {L}_1^{-1}\mathcal {L}_2$ has a sum equal to one.

Different from the previous
work \cite{li2011containment} which assumes the leaders' control inputs
$u_i$, $i\in\mathcal
{R}$, to be zero,
we consider here the general case where the leaders' control inputs
are possibly nonzero and time varying.
Suppose that the following mild
assumption holds.

{\bf Assumption 2}. The leaders' control inputs $u_i$, $i\in\mathcal
{R}$, are bounded,
i.e., $\|u_i\|\leq\gamma_i$,  $i\in\mathcal
{R}$, where $\gamma_i$ are positive
constants.

The objective of this paper is to solve the distributed containment control
problem for the agents in \dref{1c}, i.e., to design
distributed controllers under which the states of the $M$ followers
can converge to the convex hull spanned by the states
of the leaders.

Note that the leaders' control inputs are generally available to at most a subset of the followers.
In order to solve the containment problem,
based on the relative state information of neighboring agents,
we propose a distributed static controller for each follower as
\begin{equation}\label{sscdis}
\begin{aligned}
u_i &=c_1K\sum_{j=1}^Na_{ij}(x_i-x_j)
+c_2 \hat{g}\left(K\sum_{j=1}^Na_{ij}(x_i-x_j)\right),\quad i\in\mathcal {F},
\end{aligned}
\end{equation}
where
$c_1>0$ and $c_2>0\in\mathbf{R}$ are constant coupling gains,
$K\in\mathbf{R}^{p\times n}$ is the feedback gain matrix,
$a_{ij}$
is the $(i,j)$-th entry of the adjacency matrix $\mathcal {A}$
associated with $\mathcal {G}$, and the
nonlinear function $\hat{g}(\cdot)$ is defined as follows:
for $w\in\mathbf{R}^n$,
\begin{equation}\label{satu2}
\hat{g}(w)=\begin{cases}\frac{w}{\|w\|} & \text{if}~\|w\|\neq 0,\\
0 & \text{if}~\|w\|=0.
\end{cases}
\end{equation}

Let $x_f=[x_1^T,\cdots,x_M^T]^T$,
$x_l=[x_{M+1}^T,\cdots,x_{N}^T]^T$, $x=[x_f^T,x_l^T]^T$,
and $u_l=[u_{M+1}^T,\cdots,u_N^T]^T$. Then, it follows from
\dref{1c} and \dref{sscdis} that the closed-loop
network dynamics can be written as
\begin{equation}\label{netssdis}
\begin{aligned}
\dot{x}_f &= (I_M\otimes A+c_1\mathcal {L}_1\otimes BK)x_f+c_1(\mathcal {L}_2\otimes BK)x_l
+c_2(I_M\otimes B)G(x),\\
\dot{x}_l &= (I_{N-M}\otimes A)x_f+(I_M\otimes B)u_l,
\end{aligned}
\end{equation}
where $\mathcal {L}_1$ and $\mathcal {L}_2$ are defined as in \dref{lapc},
and
$\widehat{G}(x)\triangleq\begin{bmatrix}\hat{g}(K\sum_{j=1}^N a_{1j}(x_1-x_j))\\\vdots\\\hat{g}(K\sum_{j=1}^N a_{Mj}(x_M-x_j))\end{bmatrix}.$

Introduce the following variable:
\begin{equation}\label{change}
\xi\triangleq x_f+(\mathcal {L}_1^{-1}\mathcal {L}_2\otimes I_{n})x_l,
\end{equation}
where $\xi=[\xi_1^T,\cdots,\xi_M^T]^T$.
From \dref{change}, it is easy to see that
$\xi=0$ if and only if $x_f= (-
\mathcal {L}_1^{-1}\mathcal {L}_2\otimes I_n)x_l$. In virtue of
Lemma 1, we can get that the containment control problem is solved if
$\xi$ converges to zero. Hereafter, we refer to $\xi$ as the containment error.
By \dref{change} and \dref{netssdis}, it is not difficult to obtain that
$\xi$ satisfies the following dynamics:
\begin{equation}\label{netss1dis}
\begin{aligned}
\dot{\xi}
= (I_M\otimes A+c_1\mathcal {L}_1\otimes BK)\xi+c_2(I_M\otimes B)\widehat{G}(\xi)
+(\mathcal {L}_1^{-1}\mathcal {L}_2\otimes B)u_l,
\end{aligned}
\end{equation}
where
\begin{equation}\label{sa3}
\widehat{G}(\xi)=\begin{bmatrix}\hat{g}(K\sum_{j=1}^M \mathcal {L}_{1j}\xi_j)\\
\vdots\\\hat{g}(K\sum_{j=1}^M \mathcal {L}_{Mj}\xi_j)\end{bmatrix},
\end{equation}
with $\mathcal {L}_{ij}$ denoting the $(i,j)$-th entry of $\mathcal {L}_1$.

{\bf Theorem 1}. Suppose that Assumptions 1 and 2 hold.
The parameters in the containment controller \dref{sscdis}
are designed as $c_1\geq
\frac{1}{\lambda_{\min}(\mathcal
{L}_1)}$, $c_2\geq\underset{i\in\mathcal {R}}{\max}\,\gamma_i$,
and $K=-B^TP^{-1}$, where $P>0$ is a solution to the following linear matrix
inequality (LMI):
\begin{equation}\label{alg1}
AP+PA^T-2BB^T<0.
\end{equation}
Then, the containment error $\xi$ in \dref{netss1dis}
asymptotically converges to zero.

{\bf Proof.} Consider the following Lyapunov function candidate:
\begin{equation}\label{lyas1}
V_1=\frac{1}{2}\xi^T(\mathcal {L}_1\otimes P^{-1})\xi.
\end{equation}
Under Assumption 1, it follows from Lemma 1 that $\mathcal {L}_1>0$, so $V_1$ is clearly
positive definite. The time derivative of $V_1$
along the trajectory of \dref{netss1dis} is given by
\begin{equation}\label{lyas2}
\begin{aligned}
\dot{V}_1 
&=\xi^T(\mathcal {L}_1\otimes P^{-1}A+c_1\mathcal {L}_1^2\otimes P^{-1}BK)\xi\\
&\quad+c_2\xi^T(\mathcal {L}_1\otimes P^{-1}B)\widehat{G}(\xi)
+\xi^T(\mathcal {L}_2\otimes P^{-1}B)u_l\\
&=\frac{1}{2}\xi^T\mathcal {X}\xi+c_2\xi^T(\mathcal {L}_1\otimes P^{-1}B)\widehat{G}(\xi)
+\xi^T(\mathcal {L}_2\otimes P^{-1}B)u_l,
\end{aligned}
\end{equation}
where
\begin{equation}\label{x}
\mathcal {X}=\mathcal {L}_1\otimes (P^{-1}A+A^TP^{-1})-2c_1\mathcal {L}_1^2\otimes P^{-1}BB^TP^{-1}.
\end{equation}

Let $b_{ij}$ denote the $(i,j)$-th entry of $-\mathcal {L}_1^{-1}\mathcal {L}_2$,
which by Lemma 1, satisfies that
$b_{ij}\geq0$ and $\sum_{j=1}^{N-M} b_{ij}=1$ for $i=1,\cdots,M$. In virtue of
Assumption 2, we have
\begin{equation}\label{lyas3}
\begin{aligned}
\xi^T(\mathcal {L}_2\otimes P^{-1}B)u_l&=\xi^T(\mathcal {L}_1\otimes I_n)(\mathcal {L}_1^{-1}\mathcal {L}_2\otimes P^{-1}B)u_l\\
&=-\begin{bmatrix} \sum_{j=1}^M \mathcal {L}_{1j}\xi_j^T & \cdots & \sum_{j=1}^M \mathcal {L}_{Mj}\xi_j^T \end{bmatrix}
\begin{bmatrix}\sum_{k=1}^{N-M} b_{1k}P^{-1}Bu_k\\\vdots\\\sum_{k=1}^{N-M} b_{Mk}P^{-1}Bu_k\end{bmatrix}\\
&=-\sum_{i=1}^M \sum_{j=1}^M\mathcal {L}_{ij}\xi_j^TP^{-1}B\sum_{k=1}^{N-M} b_{jk}u_k\\
&\leq \sum_{i=1}^M \|B^TP^{-1}\sum_{j=1}^M\mathcal {L}_{ij}\xi_j\|\sum_{k=1}^{N-M} b_{jk}\|u_k\|\\&\leq
\sum_{i=1}^M \|B^TP^{-1}\sum_{j=1}^M\mathcal {L}_{ij}\xi_j\|\max_{i\in\mathcal {R}}\gamma_i.
\end{aligned}
\end{equation}

From \dref{satu2} and \dref{sa3},
it follows that
\begin{equation}\label{lyas4}
\begin{aligned}
\xi^T(\mathcal {L}_1\otimes P^{-1}B)\widehat{G}(\xi) &=
\begin{bmatrix} \sum_{j=1}^M\mathcal {L}_{1j}\xi_j^TP^{-1}B & \cdots & \sum_{j=1}^M\mathcal {L}_{Mj}\xi_j^TP^{-1}B \end{bmatrix}\\
&\quad\times\begin{bmatrix}-\frac{B^TP^{-1}\sum_{j=1}^M \mathcal {L}_{1j}\xi_j}{\|B^TP^{-1}\sum_{j=1}^M \mathcal {L}_{1j}\xi_j\|}\\
\vdots\\-\frac{B^TP^{-1}\sum_{j=1}^M \mathcal {L}_{Mj}\xi_j}{\|B^TP^{-1}\sum_{j=1}^M \mathcal {L}_{Mj}\xi_j\|}\end{bmatrix}\\
&=-\sum_{i=1}^M \|B^TP^{-1}\sum_{j=1}^M\mathcal {L}_{ij}\xi_j\|.
\end{aligned}
\end{equation}
Then, we can get from \dref{lyas2}, \dref{lyas3}, and \dref{lyas4} that
\begin{equation}\label{lyas5}
\begin{aligned}
\dot{V}_1 &\leq \frac{1}{2}\xi^T\mathcal {X}\xi
-(c_2-\max_{i\in\mathcal {R}}\gamma_i)\sum_{i=1}^M \|B^TP^{-1}\sum_{j=1}^M\mathcal {L}_{ij}\xi_j\|\\
&\leq\frac{1}{2}\xi^T\mathcal {X}\xi.
\end{aligned}
\end{equation}

In virtue of \dref{alg1}, we obtain that
\begin{equation}\label{lyas62}
\begin{aligned}
(\mathcal {L}_1^{-\frac{1}{2}}\otimes P)\mathcal {X}(\mathcal {L}_1^{-\frac{1}{2}}\otimes P)
&=I_M\otimes (AP+PA^T)-2c_1\mathcal {L}_1\otimes BB^T\\
&\leq I_M\otimes [AP+PA^T-2c_1\lambda_{\min}(\mathcal {L}_1)BB^T]<0,
\end{aligned}
\end{equation}
which, together with \dref{lyas5}, implies $\dot{V}_1<0$.
Therefore, the containment error $\xi$ of \dref{netss1dis} is asymptotically
stable, i.e., the containment problem is solved.
\hfill
$\blacksquare$

{\bf Remark 1}.
As shown in \cite{li2010consensus}, a necessary and
sufficient condition for the existence of a $P>0$ to the LMI
\dref{alg1} is that $(A,B)$ is stabilizable. Therefore, a sufficient
condition for the existence of \dref{sscdis} satisfying Theorem 1 is
that $(A,B)$ is stabilizable.

{\bf Remark 2}.
The distributed controller \dref{sscdis} consists of
a linear term $c_1K\sum_{j=1}^Na_{ij}(x_i-x_j)$
and a nonlinear term $c_2 \hat{g}(K\sum_{j=1}^Na_{ij}(x_i-x_j))$,
where the nonlinear term
is used to suppress the effect of
the leaders' nonzero inputs.
Without the nonlinear term in \dref{sscdis},
it can be seen from \dref{netss1dis}
that even though
$K$ is designed such that $I_M\otimes A+c_1\mathcal {L}_1\otimes BK$
is Hurwitz, the containment error will not converge
to zero due to the nonzero $u_l$.
Note that the function $\hat{g}(\cdot)$ in \dref{sscdis}
is nonsmooth, implying that the containment controller
\dref{sscdis} is discontinuous. Since the right hand of \dref{sscdis}
is measurable and locally essentially bounded,
the well-posedness and the existence of the solution to \dref{netss1dis}
can be understood in the Filippov sense \cite{shevitz1994lyapunov}.

\section{Continuous State Feedback Containment Controllers}

\subsection{Static Continuous Containment Controllers}

An inherent drawback of the discontinuous
controller \dref{sscdis}
is that
it
will result in
the undesirable chattering effect
in real implementation,
due to imperfections in switching devices
\cite{young1999control,edwards1998sliding}. 
To avoid the chattering effect, one feasible
approach is to use the boundary layer
technique \cite{young1999control,edwards1998sliding} to
give a continuous approximation of
the discontinuous function $\hat{g}(\cdot)$.

Using the boundary layer technique, we propose a distributed continuous static controller
as
\begin{equation}\label{ssc}
\begin{aligned}
u_i &=c_1K\sum_{j=1}^Na_{ij}(x_i-x_j)
+c_2 g\left(K\sum_{j=1}^Na_{ij}(x_i-x_j)\right),\quad i\in\mathcal {F},
\end{aligned}
\end{equation}
where the nonlinear function $g(\cdot)$ is defined such that
for $w\in\mathbf{R}^n$,
\begin{equation}\label{satu}
g(w)=\begin{cases}\frac{w}{\|w\|} & \text{if}~\|w\|>\kappa,\\
\frac{w}{\kappa} & \text{if}~\|w\|\leq\kappa,
\end{cases}
\end{equation}
with $\kappa$ being a small positive scalar,
denoting the width of
the boundary layer,
and the rest of the variables are
the same as in \dref{sscdis}.
It is worth mentioning that
$g(\cdot)$ is actually a saturation
function. The readers can refer
to \cite{yang2012semi,meng2013global} for previous results on
consensus
with input saturation constraints.

From \dref{change} and \dref{ssc}, we can obtain that
the containment error $\xi$ in this case
satisfies
\begin{equation}\label{netss1}
\begin{aligned}
\dot{\xi}
= (I_M\otimes A+c_1\mathcal {L}_1\otimes BK)\xi+c_2(I_M\otimes B)\widetilde{G}(\xi)
+(\mathcal {L}_1^{-1}\mathcal {L}_2\otimes B)u_l,
\end{aligned}
\end{equation}
where
\begin{equation}\label{sa4}
\widetilde{G}(\xi)\triangleq\begin{bmatrix}g(K\sum_{j=1}^M \mathcal {L}_{1j}\xi_j)\\
\vdots\\g(K\sum_{j=1}^M \mathcal {L}_{Mj}\xi_j)\end{bmatrix}.
\end{equation}

The following theorem states the ultimate boundedness of
the containment error $\xi$.

{\bf Theorem 2}. Assume that Assumptions 1 and 2 hold.
Then, the containment error $\xi$ of \dref{netss1}
under the continuous controller \dref{ssc}
with $c_1$, $c_2$, and $K$ chosen as in Theorem 1
is uniformly ultimately bounded and exponentially
converges to the residual set
\begin{equation}\label{d}
\mathcal {D}_1\triangleq \{\xi : \|\xi\|^2\leq\frac{2\lambda_{\max}(P)M\kappa\max_{i\in\mathcal {R}}\gamma_i}
{\alpha\lambda_{\min}(\mathcal {L}_1)}\},
\end{equation}
where
\begin{equation}\label{alf}
\alpha= \frac{-\lambda_{\max}(AP+PA^T-2 BB^T)}{\lambda_{\max}(P)}.
\end{equation}

{\bf Proof.} Consider the Lyapunov function $V_1$ as in the proof of Theorem 1.
The time derivative of $V_1$
along the trajectory of \dref{netss1} is given by
\begin{equation}\label{lyas2c}
\begin{aligned}
\dot{V}_1
&=\frac{1}{2}\xi^T\mathcal {X}\xi+c_2\xi^T(\mathcal {L}_1\otimes P^{-1}B)\widetilde{G}(\xi)
+\xi^T(\mathcal {L}_2\otimes P^{-1}B)u_l,
\end{aligned}
\end{equation}
where $\mathcal {X}$ is defined in \dref{x}.

Next, consider the following three cases:

i) $\|K\sum_{j=1}^M \mathcal {L}_{ij}\xi_j\|>\kappa$, i.e.,
$\|B^TP^{-1}\sum_{j=1}^M \mathcal {L}_{ij}\xi_j\|>\kappa$, $i=1,\cdots,M$.

In this case, it follows from \dref{satu} and \dref{sa4} that
\begin{equation}\label{lyas4c}
\begin{aligned}
\xi^T(\mathcal {L}_1\otimes P^{-1}B)\widetilde{G}(\xi)
=-\sum_{i=1}^M \|B^TP^{-1}\sum_{j=1}^M\mathcal {L}_{ij}\xi_j\|.
\end{aligned}
\end{equation}
Then, we can get from \dref{lyas2c}, \dref{lyas3}, and \dref{lyas4c} that
$$\begin{aligned}
\dot{V}_1 &\leq \frac{1}{2}\xi^T\mathcal {X}\xi.
\end{aligned}$$

ii) $\|B^TP^{-1}\sum_{j=1}^M \mathcal {L}_{ij}\xi_j\|\leq\kappa$, $i=1,\cdots,M$.

From \dref{lyas3}, we can obtain
that
\begin{equation}\label{lyas7c}
\xi^T(\mathcal {L}_2\otimes P^{-1}B)u_l\leq M\kappa\max_{i\in\mathcal {R}}\gamma_i.
\end{equation}
Further, it follows from \dref{satu}, \dref{sa4}, and \dref{lyas4c}
that
\begin{equation}\label{lyas8c}
\begin{aligned}
\xi^T(\mathcal {L}_1\otimes P^{-1}B)\widetilde{G}(\xi)
=-\frac{1}{\kappa}\sum_{i=1}^M \|B^TP^{-1}\sum_{j=1}^M \mathcal {L}_{ij}\xi_j\|^2
\leq 0.
\end{aligned}
\end{equation}
Thus, we get from \dref{lyas2c}, \dref{lyas7c}, and \dref{lyas8c}
that
\begin{equation}\label{lyas9c}
\begin{aligned}
\dot{V}_1 \leq \frac{1}{2}\xi^T\mathcal {X}\xi+M\kappa\max_{i\in\mathcal {R}}\gamma_i.
\end{aligned}
\end{equation}

iii) $\xi$ satisfies neither case i) nor case ii).

Without loss of generality, assume that
$\|B^TP^{-1}\sum_{j=1}^M \mathcal {L}_{ij}\xi_j\|>\kappa$, $i=1,\cdots,l$, and
$\|B^TP^{-1}\sum_{j=1}^M \mathcal {L}_{ij}\xi_j\|\leq\kappa$, $i=l+1,\cdots,M$, where $2\leq l\leq M-1$. It is easy
to see from \dref{lyas3}, \dref{satu}, and \dref{lyas4c} that
$$
\begin{aligned}
\xi^T(\mathcal {L}_2\otimes P^{-1}B)u_l &\leq
\max_{i\in\mathcal {R}}\gamma_i[\sum_{i=1}^l \|B^TP^{-1}\sum_{j=1}^M \mathcal {L}_{ij}\xi_j\|+ (M-l)\kappa],\\
\xi^T(\mathcal {L}_1\otimes P^{-1}B)\widehat{G}(\xi) &\leq-\sum_{i=1}^l \|B^TP^{-1}\sum_{j=1}^M \mathcal {L}_{ij}\xi_j\|.
\end{aligned}$$
Clearly, in this case we have
$$\begin{aligned}
\dot{V}_1 &\leq\frac{1}{2}\xi^T\mathcal {X}\xi+(M-l)\kappa\max_{i\in\mathcal {R}}\gamma_i.
\end{aligned}$$

Therefore, by analyzing the above three cases, we get that
$\dot{V}_1$ satisfies \dref{lyas9c} for all $\xi\in\mathbf{R}^{Mn}$.
Note that \dref{lyas9c} can be rewritten as
\begin{equation}\label{lyas11c}
\begin{aligned}
\dot{V}_1 &\leq -\alpha V_1+\alpha V_1+\frac{1}{2}\xi^T\mathcal {X}\xi+M\kappa\max_{i\in\mathcal {R}}\gamma_i\\
&=-\alpha V_1+\frac{1}{2}\xi^T(\mathcal {X}+\alpha\mathcal {L}_1\otimes P^{-1})\xi+M\kappa\max_{i\in\mathcal {R}}\gamma_i.
\end{aligned}
\end{equation}
Because $\alpha= \frac{-\lambda_{\max}(AP+PA^T-2 BB^T)}{\lambda_{\max}(P)}$,
in light of \dref{alg1}, we can obtain that
$$\begin{aligned}
&(\mathcal {L}_1^{-\frac{1}{2}}\otimes P)
(\mathcal {X}+\alpha\mathcal {L}_1\otimes P^{-1})
(\mathcal {L}_1^{-\frac{1}{2}}\otimes P)\\
&\quad\leq I_N\otimes [AP+PA^T+\alpha P-2BB^T]<0.
\end{aligned}$$
Then, it follows from \dref{lyas11c} that
\begin{equation}\label{lyasc13}
\begin{aligned}
\dot{V}_1
\leq-\alpha V_1+M\kappa\max_{i\in\mathcal {R}}\gamma_i.
\end{aligned}
\end{equation}
By using the Comparison lemma \cite{khalil2002nonlinear},
we can obtain from \dref{lyasc13} that
\begin{equation}\label{lyasc14}
\begin{aligned}
V_1(\xi)
\leq [V_1(\xi(0))-\frac{M\kappa}{\alpha}]{\rm{exp}(-\alpha t)}+\frac{N\kappa\max_{i\in\mathcal {R}}\gamma_i}{\alpha},
\end{aligned}
\end{equation}
which implies
that $\xi$ exponentially converges to the residual set
$\mathcal {D}_1$ in \dref{d} with a convergence rate not less than
${\rm{exp}(-\alpha t)}$.
\hfill $\blacksquare$

{\bf Remark 3}.
Contrary to the discontinuous controller \dref{sscdis},
the chattering effect can be
avoided
by using the continuous controller \dref{ssc}.
The tradeoff is that the continuous controller \dref{ssc} does not
guarantee asymptotic stability.
Note that the residual set $\mathcal {D}_1$ of the containment error $\xi$
depends on the communication graph $\mathcal {G}$,
the number of followers,
the upper bounds of the leader's control inputs,
and the width $\kappa$ of the boundary layer.
By choosing a sufficiently small $\kappa$,
$\xi$ under the continuous
controller \dref{ssc} can
be arbitrarily small, which
is acceptable in most circumstances.

\subsection{Adaptive Continuous Containment Controllers}

In the last subsection, to design the controller
\dref{ssc} we have to use the minimal eigenvalue $\lambda_{\min}(\mathcal {L}_1)$ of
$\mathcal {L}_1$ and the upper bounds $\gamma_i$ of the leaders'
control inputs. However,
$\lambda_{\min}(\mathcal {L}_1)$ is global information in the sense that
each follower has to know the entire
communication graph to compute it
and
it is not practical
to assume that
the upper bounds $\gamma_i$, $i\in\mathcal {R}$,
are explicitly known
to all followers.
In this subsection, we intend to
design distributed controllers
to solve the containment problem
without requiring
$\lambda_{\min}(\mathcal {L}_1)$ nor $\gamma_i$, $i\in\mathcal {R}$.

Based on the relative states of neighboring agents, we
propose the following distributed controller with an adaptive law
for updating the coupling gain for each follower:
\begin{equation}\label{acons}
\begin{aligned}
u_i&=d_iK\sum_{j=1}^Na_{ij}(x_i-x_j)+d_i(t)r\left(K\sum_{j=1}^Na_{ij}(x_i-x_j)\right),\\
\dot{d}_i &=\tau_i\left(-\varphi_id_i+[\sum_{j=1}^Na_{ij}(x_i-x_j)^T]
\Gamma[\sum_{j=1}^Na_{ij}(x_i-x_j)]\right.\\
&\quad +\left.\|K\sum_{j=1}^Na_{ij}(x_i-x_j)\|\right),\quad i=1,\cdots,M,
\end{aligned}
\end{equation}
where $d_i(t)$ denotes the
time-varying coupling gain associated with the $i$-th follower,
$\varphi_i$ are small positive constants,
$\Gamma\in\mathbf{R}^{n\times n}$ is the feedback gain
matrix, $\tau_i$ are positive scalars, the nonlinear function $r(\cdot)$
is defined as follows: for $w\in\mathbf{R}^n$,
\begin{equation}\label{satua}
r(w)=\begin{cases} \frac{w}{\|w\|} & \text{if}~d_i\|w\|>\kappa,\\
\frac{w}{\kappa}d_i & \text{if}~d_i\|w\|\leq\kappa,
\end{cases}
\end{equation}
and the rest of the variables
are defined as in \dref{ssc}.

Let $x_f$,
$x_l$, $x$, $u_l$, and $\xi$ be defined as in \dref{netss1} and \dref{change}.
Let $D(t)=\rm{diag}(d_1(t),\cdots,d_M(t))$.
Then, it follows from
\dref{1c} and \dref{acons} that the containment error
$\xi$ and the coupling
gains $D(t)$ satisfy the following dynamics:
\begin{equation}\label{netas1}
\begin{aligned}
\dot{\xi}
&= \dot{x}_f+(\mathcal {L}_1^{-1}\mathcal {L}_2\otimes I_{n})\dot{x}_l\\
&= (I\otimes A+D\mathcal
{L}_1\otimes BK)x_f+(D\mathcal
{L}_2\otimes BK)x_l+(D\otimes B)R(x)\\&\quad+(\mathcal {L}_1^{-1}\mathcal {L}_2\otimes A)x_l+(\mathcal {L}_1^{-1}\mathcal {L}_2\otimes B)u_l\\
&= (I_M\otimes A+D\mathcal {L}_1\otimes BK)\xi+(D\otimes B)R(\xi)+(\mathcal {L}_1^{-1}\mathcal {L}_2\otimes B)u_l,\\
\dot{d}_i &=\tau_i\left(-\varphi_id_i+[\sum_{j=1}^M\mathcal {L}_{ij}\xi_j^T]\Gamma[\sum_{j=1}^M\mathcal {L}_{ij}\xi_j]
+\|K\sum_{j=1}^M\mathcal {L}_{ij}\xi_j\|\right),\quad i=1,\cdots,M,
\end{aligned}
\end{equation}
where
$R(\xi)=\begin{bmatrix}r(K\sum_{j=1}^M \mathcal {L}_{1j}\xi_j)\\
\vdots\\r(K\sum_{j=1}^M \mathcal {L}_{Mj}\xi_j)\end{bmatrix}.$

The following theorem shows the ultimate boundedness of
the states $\xi$ and $d_i$ of \dref{netas1}.

{\bf Theorem 3}. Suppose that Assumptions 1 and 2 hold.
The feedback gain matrices of the adaptive controller \dref{acons}
are designed as
$K=-B^TP^{-1}$ and $\Gamma=P^{-1}BB^TP^{-1}$, where
$P>0$ is a solution to the LMI \dref{alg1}.
Then, both the containment error $\xi$
and the coupling gains $d_i$, $i=1,\cdots,M$, in \dref{netas1}
are uniformly ultimately bounded.
Furthermore, if $\varphi_i$ and $\psi_i$ are chosen such that
$\varrho\triangleq\max_{i=1,\cdots,M}\varphi_i \tau_i<\alpha$,
where $\alpha$ is defined as in \dref{alf},
then $\xi$ exponentially converges to the residual set
\begin{equation}\label{d2k}
\mathcal {D}_2\triangleq \{\xi:\|\xi\|^2\leq
\frac{\lambda_{\max}(P)}{\lambda_2(\alpha-\varrho)}[\sum_{i=1}^N\beta^2\varphi_i+\frac{1}{2}M\kappa]\},
\end{equation}
where $\beta\geq \underset{i\in\mathcal {R}}{\max}\{\,\gamma_i,\frac{1}{\lambda_{\min}(\mathcal {L}_1)}\}$.

{\bf Proof}. Let $\tilde{d}_i=d_i-\beta$, $i=1,\cdots,M$. Then, \dref{netas1}
can be rewritten as
\begin{equation}\label{netas2}
\begin{aligned}
\dot{\xi}
&= (I_M\otimes A+\widetilde{D}\mathcal {L}_1\otimes BK)\xi+(\widetilde{D}\otimes B)R(\xi)+(\mathcal {L}_1^{-1}\mathcal {L}_2\otimes B)u_l,\\
\dot{\tilde{d}}_i &=\tau_i\left(-\varphi_i(\tilde{d}_i+\beta)+[\sum_{j=1}^M\mathcal {L}_{ij}\xi_j^T]\Gamma[\sum_{j=1}^M\mathcal {L}_{ij}\xi_j]
+\|K\sum_{j=1}^M\mathcal {L}_{ij}\xi_j\|\right),\quad i=1,\cdots,M,
\end{aligned}
\end{equation}
where $\widetilde{D}(t)=\rm{diag}(\tilde{d}_1(t)+\beta,\cdots,\tilde{d}_M(t)+\beta)$.

Consider the following Lyapunov function candidate
$$V_2=\frac{1}{2}\xi^T(\mathcal {L}_1\otimes P^{-1})\xi+\sum_{i=1}^M\frac{\tilde{d}_i^2}{2\tau_i},
$$
As stated in the proof of Theorem 1,
it is easy to see that $V_2$ is positive definite. The time derivative of $V_2$ along \dref{netas2}
can be obtained as
\begin{equation}\label{lyaas2}
\begin{aligned}
\dot{V}_2
&= \xi^T(\mathcal {L}_1\otimes P^{-1}A+\mathcal {L}_1\widetilde{D}\mathcal {L}_1\otimes P^{-1}BK)\xi\\
&\quad +\xi^T(\mathcal {L}_1\widetilde{D}\otimes P^{-1}B)R(\xi)+2\xi^T(\mathcal {L}_2\otimes P^{-1}B)u_l\\
&\quad+\sum_{i=1}^M\tilde{d}_i\left(-\varphi_i(\tilde{d}_i+\beta)+[\sum_{j=1}^M\mathcal {L}_{ij}\xi_j^T]\Gamma[\sum_{j=1}^M\mathcal {L}_{ij}\xi_j]
+\|K\sum_{j=1}^M\mathcal {L}_{ij}\xi_j\|\right).
\end{aligned}
\end{equation}

By substituting $K=-BP^{-1}$, it is easy to get that
\begin{equation}\label{lyaas3}
\xi^T(\mathcal {L}_1\widetilde{D}\mathcal {L}_1\otimes P^{-1}BK)\xi=
-\sum_{i=1}^M(\tilde{d}_i+\beta)[\sum_{j=1}^M\mathcal {L}_{ij}\xi_j]^TP^{-1}BB^TP^{-1}[\sum_{j=1}^M\mathcal {L}_{ij}\xi_j].
\end{equation}

For the case where $d_i\|K\sum_{j=1}^M\mathcal {L}_{ij}\xi_j\|>\kappa$, $i=1,\cdots,M$,
we can get from \dref{satua} that
\begin{equation}\label{lyaas4}
\begin{aligned}
&\xi^T(\mathcal {L}_1\tilde{D}\otimes P^{-1}B)R(\xi)=-\sum_{i=1}^M(\tilde{d}_i+\beta)\|B^TP^{-1}\sum_{j=1}^M \mathcal {L}_{ij}\xi_j\|.
\end{aligned}
\end{equation}
Substituting \dref{lyaas3}, \dref{lyaas4}, and \dref{lyas3} into \dref{lyaas2} yields
$$\begin{aligned}
\dot{V}_2
&\leq \xi^T(\mathcal {L}_1\otimes P^{-1}A)\xi-
\beta \sum_{i=1}^M [\sum_{j=1}^M\mathcal {L}_{ij}\xi_j]^TP^{-1}BB^TP^{-1}[\sum_{j=1}^M\mathcal {L}_{ij}\xi_j]\\
&\quad -(\beta-\max_{i\in\mathcal {R}}\gamma_i)\sum_{i=1}^M\|B^TP^{-1}\sum_{j=1}^M \mathcal {L}_{ij}\xi_j\|
-\sum_{i=1}^M\varphi_i(\tilde{d}_i^2+\tilde{d}_i\beta)\\
&\leq \frac{1}{2}\xi^T\mathcal {Z}\xi
+\frac{1}{2}\sum_{i=1}^M\varphi_i(-\tilde{d}_i^2+\beta^2),
\end{aligned}$$
where
we have used the fact that $\beta\geq\underset{i\in\mathcal {R}}{\max}\,\gamma_i$ and $-\tilde{d}_i^2-\tilde{d}_i\beta\leq-\frac{1}{2}\tilde{d}_i^2+\frac{1}{2}\beta^2$
to get the last inequality and
$$\mathcal {Z}=\mathcal {L}_1\otimes (P^{-1}A+A^TP^{-1})-2\beta \mathcal {L}_1^2\otimes P^{-1}BBP^{-1}.$$

For the case where $d_i\|K\sum_{j=1}^M\mathcal {L}_{ij}\xi_j\|\leq\kappa$, $i=1,\cdots,M$,
we can get from \dref{satua} that
\begin{equation}\label{lyaas6}
\begin{aligned}
&\xi^T(\mathcal {L}_1\tilde{D}\otimes P^{-1}B)R(\xi)=-\sum_{i=1}^M\frac{(\tilde{d}_i+\beta)^2}{\kappa}\|B^TP^{-1}\sum_{j=1}^M \mathcal {L}_{ij}\xi_j\|.
\end{aligned}
\end{equation}
Then, it follows from \dref{lyaas3}, \dref{lyaas6}, \dref{lyas3}, and \dref{lyaas2} that
\begin{equation}\label{lyaas7}
\begin{aligned}
\dot{V}_2
&\leq \frac{1}{2}\xi^T\mathcal {Z}\xi
+\sum_{i=1}^M\varphi_i(\tilde{d}_i^2+\tilde{d}_i\beta)\\
&\quad-\sum_{i=1}^M\frac{(\tilde{d}_i+\beta)^2}{\kappa}\|B^TP^{-1}\sum_{j=1}^M \mathcal {L}_{ij}\xi_j\|^2
+\sum_{i=1}^M(\tilde{d}_i+\beta)\|B^TP^{-1}\sum_{j=1}^M \mathcal {L}_{ij}\xi_j\|\\
&\leq \frac{1}{2}\xi^T\mathcal {Z}\xi
+\frac{1}{2}\sum_{i=1}^M\varphi_i(-\tilde{d}_i^2+\beta^2)+\frac{1}{4}M\kappa.
\end{aligned}
\end{equation}
Note that to get the last inequality in \dref{lyaas7}, we have used the following fact:
$$-\frac{(\tilde{d}_i+\beta)^2}{\kappa}\|B^TP^{-1}\sum_{j=1}^M \mathcal {L}_{ij}\xi_j\|^2
+(\tilde{d}_i+\beta)\|B^TP^{-1}\sum_{j=1}^M \mathcal {L}_{ij}\xi_j\|\leq \frac{1}{4}\kappa,$$
for $(\tilde{d}_i+\beta)\|K\sum_{j=1}^M\mathcal {L}_{ij}\xi_j\|\leq\kappa$, $i=1,\cdots,M$.

For the case where $d_i\|K\sum_{j=1}^M\mathcal {L}_{ij}\xi_j\|\leq\kappa$, $i=1,\cdots,l$, and
$d_i\|K\sum_{j=1}^M\mathcal {L}_{ij}\xi_j\|>\kappa$, $i=l+1,\cdots,M$. By following the
steps in the two cases above, it is easy to get that
$$\begin{aligned}
\dot{V}_2
&\leq \frac{1}{2}\xi^T\mathcal {Z}\xi
+\frac{1}{2}\sum_{i=1}^M\varphi_i(-\tilde{d}_i^2+\beta^2)+\frac{1}{4}(M-l)\kappa.
\end{aligned}$$

Therefore, $\dot{V}_2$ satisfies \dref{lyaas7} for all $\xi\in\mathbf{R}^{Nn}$.
Because $\beta\lambda_{\min}(\mathcal {L}_1)\geq1$,
by following similar steps as in the proof of Theorem 1,
it is easy to shown that $\mathcal {Z}<0$ and thereby
$\xi^T\mathcal {Z}\xi-\sum_{i=1}^M\varphi_i\tilde{d}_i^2<0$.
In virtue of the result in \cite{corless1981continuous},
we get that the states $\xi$
and $d_i$ of \dref{netas1}
are uniformly ultimately bounded.

Next, we will derive the residual set for the containment error $\xi$.
Rewrite \dref{lyaas7} into
\begin{equation}\label{lyasac71}
\begin{aligned}
\dot{V}_2
&\leq
-\varrho V_2+\frac{1}{2}\xi^T(\mathcal {Z}+\alpha\mathcal {L}_1\otimes P^{-1})\xi-\frac{1}{2}\sum_{i=1}^M(\varphi_i-\frac{\varrho}{\tau_i})\tilde{d}_i^2\\
&\quad-\frac{\alpha-\varrho}{2}\xi^T(\mathcal {L}_1\otimes P^{-1})\xi
+\frac{1}{2}\sum_{i=1}^M\beta^2\varphi_i+\frac{1}{4}M\kappa\\
&\leq -\varrho V_2-\frac{\lambda_{\min}(\mathcal {L}_1)(\alpha-\varrho)}{2\lambda_{\max}(P)}\|\xi\|^2
+\frac{1}{2}\sum_{i=1}^M\beta^2\varphi_i+\frac{1}{4}M\kappa.
\end{aligned}
\end{equation}
Obviously, it follows from \dref{lyasac71} that $\dot{V}_2\leq -\varrho V_2$ if
$\|\xi\|^2>\frac{\lambda_{\max}(P)}{\lambda_{\min}(\mathcal {L}_1)(\alpha-\varrho)}[\sum_{i=1}^M \beta^2\varphi_i+\frac{1}{2}M\kappa].$
Then, by noting $V_2\geq \frac{\lambda_{\min}(\mathcal {L}_1)}{2\lambda_{\max}(P)}\|\xi\|^2$,
we can get that if $\varrho\leq\alpha$ then
$\xi$ exponentially converges to the residual set $\mathcal {D}_2$ in \dref{d2k}
with a convergence rate faster than ${\rm{exp}(-\varrho t)}$.
\hfill $\blacksquare$

{\bf Remark 4}.
It is worth mentioning that introducing the term $-\varphi_id_i$ into \dref{acons}
is inspiring the $\sigma$-modification technique in the classic adaptive literature \cite{ioannou1984instability},
which plays a vital role to guarantee the ultimate
boundedness of the containment error $\xi$
and the adaptive gains $d_i$.
From \dref{d2k}, we can observe that
the residual set $\mathcal {D}_2$
decrease as $\kappa$ and $\varphi_i$ decrease.
Therefore, we can choose $\varphi_i$
and $\kappa$ to be relatively small in order to guarantee
a small containment error $\xi$.
Contrary to the fixed containment controller \dref{ssc},
the design of the
adaptive controller \dref{acons}
relies on only the agent dynamics,
requiring neither the minimal
eigenvalue $\lambda_1(\mathcal {L}_1)$ nor the upper bounds
of the leaders' control input.

\section{Continuous Output Feedback Containment Controllers}

The containment controllers in the proceeding sections are
based on the relative state information of neighboring agents,
which might not be available in some circumstances.
In this section, we extend to consider
the case where the outputs, rather not the states, of the agents are accessible to their
local neighbors.

To achieve containment,
we propose the following distributed observer-based containment controller
with fixed coupling gains:
\begin{equation}\label{sdc}
\begin{aligned}
\dot{v}_i &=Av_i+Bu_i+L(Cv_i-y_i),\\
u_i &=c_1F\sum_{j=1}^Na_{ij}(v_i-v_j)+c_2g\left(F\sum_{j=1}^Na_{ij}(v_i-v_j)\right),\quad i\in\mathcal {F},
\end{aligned}
\end{equation}
where
$v_i\in\mathbf{R}^{n}$ is the estimate of the state of
the $i$-th follower, $v_j\in\mathbf{R}^{n}$ denotes the
estimate of the state of the $j$-th leader, given by
\begin{equation}\label{observer}
\dot{v}_j=Av_j+Bu_j+L(Cv_j-y_j),\quad j\in\mathcal {R},
\end{equation}
$L\in\mathbf{R}^{q\times n}$ and
$F\in\mathbf{R}^{p\times n}$ are the feedback gain matrices,
and the rest of the variables are defined as in \dref{ssc}.
Distributed observer-based containment controllers
with adaptive coupling gains can be similarly given, which are omitted here
for brevity.

Let $z_i=[x_i^T,v_i^T]^T$, $z_f=[z_1^T,\cdots,z_M^T]^T$,
$z_l=[z_{M+1}^T,\cdots,z_{N}^T]^T$, and $z=[z_f^T,z_l^T]^T$. Then, the closed-loop network
dynamics resulting from \dref{1c} and \dref{sdc} can be written as
\begin{equation}\label{netdd}
\begin{aligned}
\dot{z}_f &= \left(I_M\otimes \mathcal {M}+c\mathcal
{L}_1\otimes\mathcal {H}\right)z_f+c_1(\mathcal {L}_2\otimes\mathcal {H})z_l
+c_2(I_M\otimes \mathcal {B})H(z),\\
\dot{z}_l &= \left(I_{N-M}\otimes \mathcal {M}\right)z_l+(I_{N-M}\otimes \mathcal {B})u_l,
\end{aligned}
\end{equation}
where
$$\begin{aligned}
\mathcal {M} &=\begin{bmatrix}A & 0\\-LC &
A+LC\end{bmatrix}, \quad \mathcal {H}=\begin{bmatrix}0 & BF\\
0 & BF\end{bmatrix},
\quad\mathcal {B}=\begin{bmatrix}B\\
B\end{bmatrix},\\
H(z)&\triangleq\begin{bmatrix} g(\mathcal {J}\sum_{j=1}^N a_{1j}(z_1-z_j))\\
\vdots\\ g(\mathcal {J}\sum_{j=1}^N a_{Mj}(z_M-z_j))\end{bmatrix}, \quad \mathcal {J}=\begin{bmatrix}0 & F
\end{bmatrix}.
\end{aligned}$$
Introduce the containment error in this case as
\begin{equation}\label{change2}
\zeta=z_f+(\mathcal {L}_1^{-1}\mathcal {L}_2\otimes I_{2n})z_l,
\end{equation}
where $\zeta=[\zeta_1^T,\cdots,\zeta_M^T]^T$.
Similarly to the proceeding section, it is easy to get that
$\zeta$ satisfies
\begin{equation}\label{netdd2}
\begin{aligned}
\dot{\zeta} &= \left(I_M\otimes \mathcal {M}+c_1\mathcal
{L}_1\otimes\mathcal {H}\right)\zeta+c_2(I_M\otimes \mathcal {S})
H(\zeta)
+(\mathcal {L}_1^{-1}\mathcal
{L}_2\otimes \mathcal {B})u_l,
\end{aligned}
\end{equation}
where
$H(\zeta)=\begin{bmatrix} g(\mathcal {J}\sum_{j=1}^M \mathcal {L}_{1j}\zeta_j)\\
\vdots\\ g(\mathcal {J}\sum_{j=1}^M \mathcal {L}_{Mj}\zeta_j)\end{bmatrix}.$

{\bf Theorem 4}. Suppose that Assumptions 1 and 2 hold.
Design the parameters of the observer-based controller \dref{sdc}
such that $A+LC$ is Hurwitz, $c_1\geq
\frac{1}{\lambda_{\min}(\mathcal {L}_1)}$, $c_2\geq\underset{i\in\mathcal {R}}{\max}\,\gamma_i$, and $K=-B^TP^{-1}$, where
$P>0$ is a solution to the LMI \dref{alg1}. The containment
error $\zeta$ described by \dref{netdd2} is uniformly ultimately
bounded.

{\bf Proof}. Consider the following Lyapunov function candidate
$$V_3=\frac{1}{2}\zeta^T(\mathcal {L}_1\otimes \mathcal {Q})\zeta,$$
where $\mathcal {Q}\triangleq\begin{bmatrix}\varsigma Q & -\varsigma Q\\
-\varsigma Q & \varsigma Q+P^{-1}\end{bmatrix}$,
$Q>0$ satisfies that $(A+LC)Q+(A+LC)^TQ<0$, and $\varsigma>0$ is a positive scalar
to be determined later.
By Schur Complement Lemma \cite{boyd1994linear}, it is easy to verify that $\mathcal {Q}>0$.
Because $\mathcal {L}_1>0$, $V_3$ is
positive definite.
The time derivative of $V_3$ along \dref{netdd2} can be obtained as
\begin{equation}\label{lyasd2}
\begin{aligned}
\dot{V}_3 
&=\zeta^T(\mathcal {L}_1\otimes\mathcal {Q}\mathcal {M}
+c_1\mathcal {L}_1^2\otimes \mathcal {Q}\mathcal {H})\zeta\\
&\quad+c_2\zeta^T(\mathcal {L}_1\otimes \mathcal {Q}\mathcal {B})\widehat{H}(\zeta)
+\zeta^T(\mathcal {L}_2\otimes
\mathcal {Q}\mathcal {B})u_l.
\end{aligned}
\end{equation}
Let $\tilde{\zeta}=(I_M\otimes T)\zeta$ with $T=\begin{bmatrix}I & -I\\
0 & I\end{bmatrix}$.
Then, \dref{lyasd2} can be rewritten as
\begin{equation}\label{lyasd21}
\begin{aligned}
\dot{V}_3 &=\frac{1}{2}\tilde{\zeta}^T \mathcal {Y}\tilde{\zeta}
+c_2\tilde{\zeta}^T(\mathcal {L}_1\otimes \widetilde{\mathcal {Q}}\widetilde{\mathcal {B}})
\widehat{H}(\tilde{\zeta}) +\tilde{\zeta}^T(\mathcal {L}_2\otimes
\widetilde{\mathcal {Q}}\widetilde{\mathcal {B}})u_l,
\end{aligned}
\end{equation}
where we have used the fact that $\mathcal {J}T^{-1}=\mathcal {J}$ and
$$\begin{aligned}
\mathcal {Y} &\triangleq\mathcal {L}_1\otimes(\widetilde{\mathcal {Q}}\widetilde{\mathcal {M}}+\widetilde{\mathcal {M}}^T\widetilde{\mathcal {Q}})
+2c_1\mathcal {L}_1^2\otimes \widetilde{\mathcal {Q}}\widetilde{\mathcal {H}},\\
\widetilde{\mathcal {Q}}& =\begin{bmatrix} \varsigma Q & 0\\
0 & P^{-1}\end{bmatrix},~\widetilde{\mathcal {M}} =\begin{bmatrix}A+LC & 0\\
-LC & A\end{bmatrix},~
\widetilde{\mathcal {H}}  =\begin{bmatrix}0 & 0\\
0 & BF\end{bmatrix},
~\widetilde{\mathcal {B}}=\begin{bmatrix}0\\
B\end{bmatrix}.
\end{aligned}$$

Consider the case where
$\|B^TP^{-1}\sum_{j=1}^M \mathcal {L}_{ij}\zeta_j\|>\kappa$, $i=1,\cdots,M$.
By noting that
$
\mathcal {J}=-\begin{bmatrix}0 & B^TP^{-1}\end{bmatrix}=
-\widetilde{\mathcal {B}}^T\widetilde{\mathcal {Q}},$
it is not difficult to get that
\begin{equation}\label{lnot}
\begin{aligned}
&\tilde{\zeta}^T(\mathcal {L}_1\otimes \widetilde{\mathcal {Q}}\widetilde{\mathcal {B}})
\widehat{H}(\tilde{\zeta})=-\sum_{i=1}^M \|\widetilde{\mathcal {B}}^T\widetilde{\mathcal {Q}}\sum_{j=1}^M\mathcal {L}_{ij}\tilde{\zeta}_j\|.
\end{aligned}
\end{equation}
By following the similar steps in \dref{lyas3}, we can get that
\begin{equation}\label{lyasd31}
\begin{aligned}
\tilde{\zeta}^T(\mathcal {L}_2\otimes
\widetilde{\mathcal {Q}}\widetilde{\mathcal {B}})u_l\leq
\max_{i\in\mathcal {R}}\gamma_i\sum_{i=1}^M \|\widetilde{\mathcal {B}}^T\widetilde{\mathcal {Q}}
\sum_{j=1}^M\mathcal {L}_{ij}\tilde{\zeta}_j\|.
\end{aligned}
\end{equation}
Substituting
\dref{lnot} and \dref{lyasd31} into \dref{lyasd21} yields
$$\begin{aligned}
\dot{V}_3 &\leq
\frac{1}{2}\tilde{\zeta}^T\mathcal {Y}\tilde{\zeta}-(c_2-\max_{i\in\mathcal {R}}\gamma_i)
\sum_{i=1}^M \|\widetilde{\mathcal {B}}^T\widetilde{\mathcal {Q}}\sum_{j=1}^M\mathcal {L}_{ij}\tilde{\zeta}_j\|\\
&\leq \frac{1}{2}\tilde{\zeta}^T\mathcal {Y}\tilde{\zeta}.
\end{aligned}$$

For the case where $\|B^TP^{-1}\sum_{j=1}^M \mathcal {L}_{ij}\tilde{\zeta}_j\|\leq\kappa$, $i=1,\cdots,M$,
it is easy
to see from \dref{lyasd31}, \dref{satu}, and \dref{lnot} that
$$
\begin{aligned}
\tilde{\zeta}^T(\mathcal {L}_1\otimes \widetilde{\mathcal {Q}}\widetilde{\mathcal {B}})\widehat{H}(\zeta) &\leq-\frac{1}{\kappa}\sum_{i=1}^M
\|\widetilde{\mathcal {B}}^T\widetilde{\mathcal {Q}}\sum_{j=1}^M\mathcal {L}_{ij}\tilde{\zeta}_j\|^2\leq0,\\
\tilde{\zeta}^T(\mathcal {L}_2\otimes
\widetilde{\mathcal {Q}}\widetilde{\mathcal {B}})u_l&\leq
M\kappa\max_{i\in\mathcal {R}}\gamma_i.
\end{aligned}
$$
Clearly, in this case we have
\begin{equation}\label{lyasd11}
\begin{aligned}
\dot{V}_3 &\leq\frac{1}{2}\tilde{\zeta}^T\mathcal {Y}\tilde{\zeta}+M\kappa \max_{i\in\mathcal {R}}\gamma_i.
\end{aligned}
\end{equation}

For the case where $\|B^TP^{-1}\sum_{j=1}^M \mathcal {L}_{ij}\tilde{\zeta}_j\|>\kappa$, $i=1,\cdots,l$, and
$\|B^TP^{-1}\sum_{j=1}^M \mathcal {L}_{ij}\tilde{\zeta}_j\|\leq\kappa$, $i=l+1,\cdots,M$,
by following similar steps in the above two cases, it is not difficult to get that
$$\begin{aligned}
\dot{V}_3 \leq\frac{1}{2}\tilde{\zeta}^T\mathcal {Y}\tilde{\zeta}+(M-l)\kappa\max_{i\in\mathcal {R}}\gamma_i.
\end{aligned}$$

Therefore, we obtain from the above three cases that $\dot{V}_3$ satisfies \dref{lyasd11}
for all $\zeta\in\mathbf{R}^{2Nn}$. By noting that $\widetilde{\mathcal {Q}}\widetilde{\mathcal {H}}\leq0$,
we have
\begin{equation}\label{lyasd4}
\begin{aligned}
(\mathcal {L}_1^{-\frac{1}{2}}\otimes I_{2n})\mathcal {Y}(\mathcal {L}_1^{-\frac{1}{2}}\otimes I_{2n})&
\leq I_M\otimes[\widetilde{\mathcal {Q}}\widetilde{\mathcal {M}}
+\widetilde{\mathcal {M}}^T\widetilde{\mathcal {Q}}+2c_1\lambda_{\min}(\mathcal {L}_1)\widetilde{\mathcal {Q}}\widetilde{\mathcal {H}}].
\end{aligned}
\end{equation}
Furthermore,
\begin{equation}\label{lyaet25}
\begin{aligned}
&\rm{diag}(I,P)[\widetilde{\mathcal {Q}}\widetilde{\mathcal {M}}+\widetilde{\mathcal {M}}^T\widetilde{\mathcal {Q}}
+2c_1\lambda_{\min}(\mathcal {L}_1)\widetilde{\mathcal {Q}}\widetilde{\mathcal {H}}]\rm{diag}(I,P)\\
&\quad=\begin{bmatrix}\varsigma[Q(A+LC)+(A+LC)^TQ] & -C^TL^T\\
-LC & AP+PA^T-2c_1\lambda_{\min}(\mathcal {L}_1)BB^T\end{bmatrix}.
\end{aligned}
\end{equation}
Because $c_1\lambda_{\min}(\mathcal {L}_1)\geq1$, it follows from
\dref{alg1} that
$AP+PA^T-2c_1\lambda_{\min}(\mathcal {L}_1)BB^T<0$.
Then, by choosing $\varsigma>0$ sufficiently large
and using Schur Complement Lemma \cite{boyd1994linear},
we can obtain that $
\widetilde{\mathcal {Q}}\widetilde{\mathcal {M}}+\widetilde{\mathcal {M}}^T\widetilde{\mathcal {Q}}
+2c_1\lambda_{\min}(\mathcal {L}_1)\widetilde{\mathcal {Q}}\widetilde{\mathcal {H}}<0$. Then, it
follows from \dref{lyasd4} and \dref{lyaet25}
that $\mathcal {Y}<0$.
Therefore, we get from \dref{lyasd11} that
the containment error $\zeta$ is uniformly
ultimately bounded.
\hfill
$\blacksquare$

{\bf Remark 5}. Containment control of multi-agent systems
was previously studied in
\cite{ji2008containment,cao2009containment,cao2011distributed2,cao2012distributed,lou2012target,
mei2012distributed,dimarogonas2009leader,meng2010distributed,li2011containment}.
The agent dynamics are restricted to be single or double
integrators in
\cite{ji2008containment,cao2009containment,cao2011distributed2,cao2012distributed,lou2012target} and
to be second-order Euler-Lagrange systems in
\cite{mei2012distributed,dimarogonas2009leader,meng2010distributed}.
In \cite{li2011containment}, it is assumed that
the leaders' control inputs are zero. In contrast,
Theorems 1-4 obtained in this paper
are applicable to multi-agent systems with general linear
dynamics and
multiple leaders whose control inputs
are possibly nonzero and bounded.
Furthermore,
contrary to the discontinuous
controllers in \cite{cao2009containment,cao2011distributed2,cao2012distributed,meng2010distributed,
mei2012distributed},
a distinct feature
of the proposed containment
controllers \dref{ssc},
\dref{acons}, and \dref{sdc}
is that they are continuous,
and thus the undesirable chattering
phenomenon can be avoided.
Another contribution of this paper
is that the adaptive containment controller \dref{acons}
can be implemented in a fully distributed
fashion without requiring
any global information.

\section{Simulation Examples}

In this section,
a simulation example is provided to validate the
effectiveness of the theoretical results.

\begin{figure}[htbp]
\centering
\includegraphics[width=0.4\linewidth]{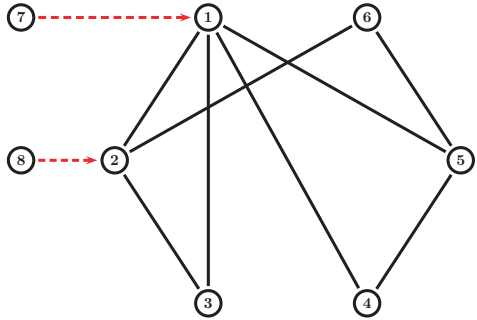}
\caption{The communication graph. }
\end{figure}

Consider a network
of eight agents with matching uncertainties.
For illustration, let the communication
graph among the agents
be given as in Figure 1, where nodes 7 and 8 are
two leaders and the others are followers.
The dynamics of the agents are given by \dref{1c}, with
$$\begin{aligned}
x_i=\begin{bmatrix}x_{i1}\\x_{i2} \end{bmatrix},\quad
A=\begin{bmatrix}0 & 1\\ -1 &1\end{bmatrix},
\quad B=\begin{bmatrix}0 \\ 1\end{bmatrix},
\end{aligned}$$
Design the control inputs for the leaders as
$u_7=K_7x_7+4\sin(2t)$ and $u_8=K_8x_8+2\cos(t)$,
with $K_7=-\begin{bmatrix}0 &
2\end{bmatrix}$
and $K_8=-\begin{bmatrix} 1 &
3\end{bmatrix}$.
It is easy to see that in this case $u_7$ and $u_8$ are
bounded. Here we use the adaptive control
\dref{acons} to solve the containment control problem.

\begin{figure}[htbp]\centering
\includegraphics[width=0.45\linewidth,height=0.3\linewidth]{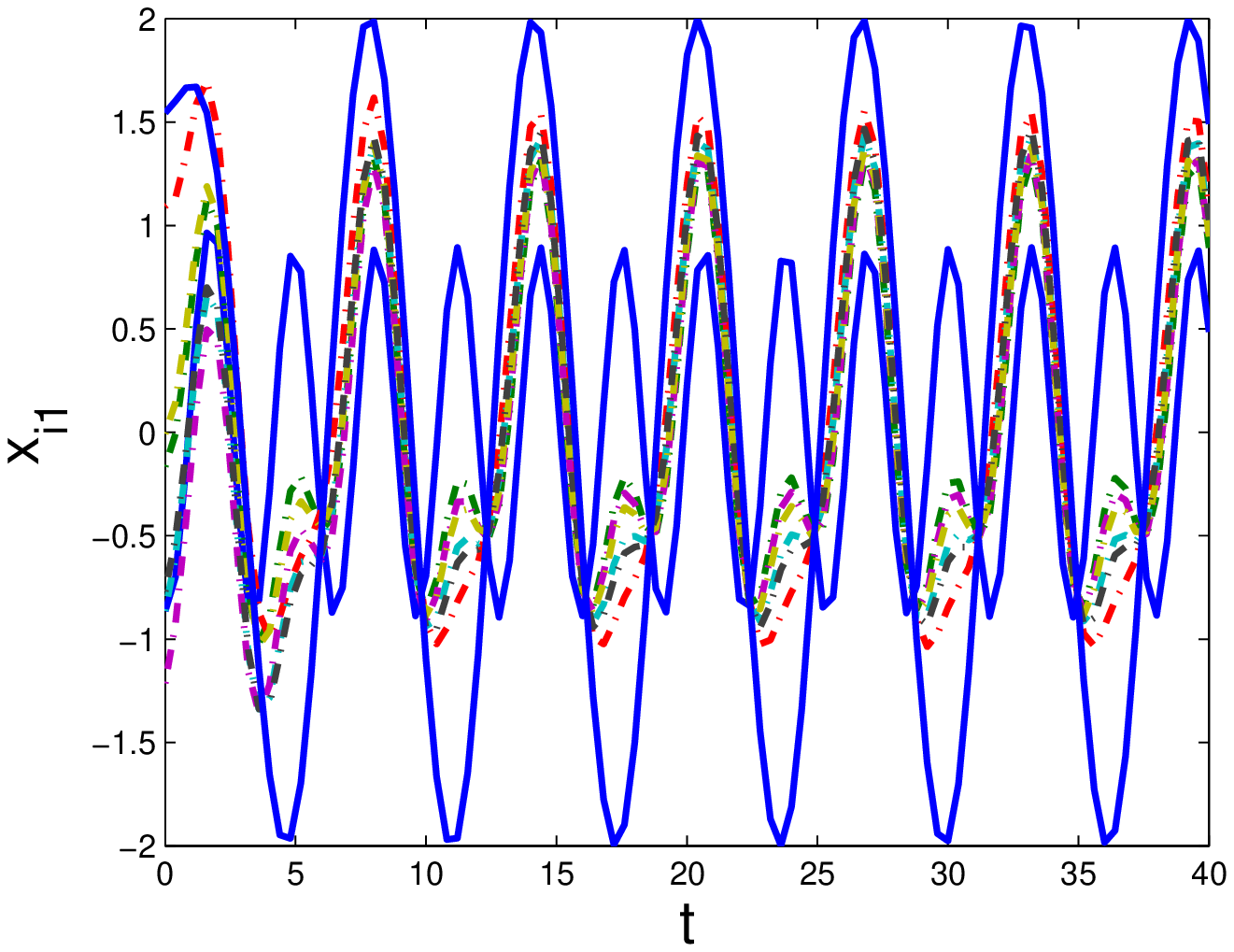}~
\includegraphics[width=0.45\linewidth,height=0.3\linewidth]{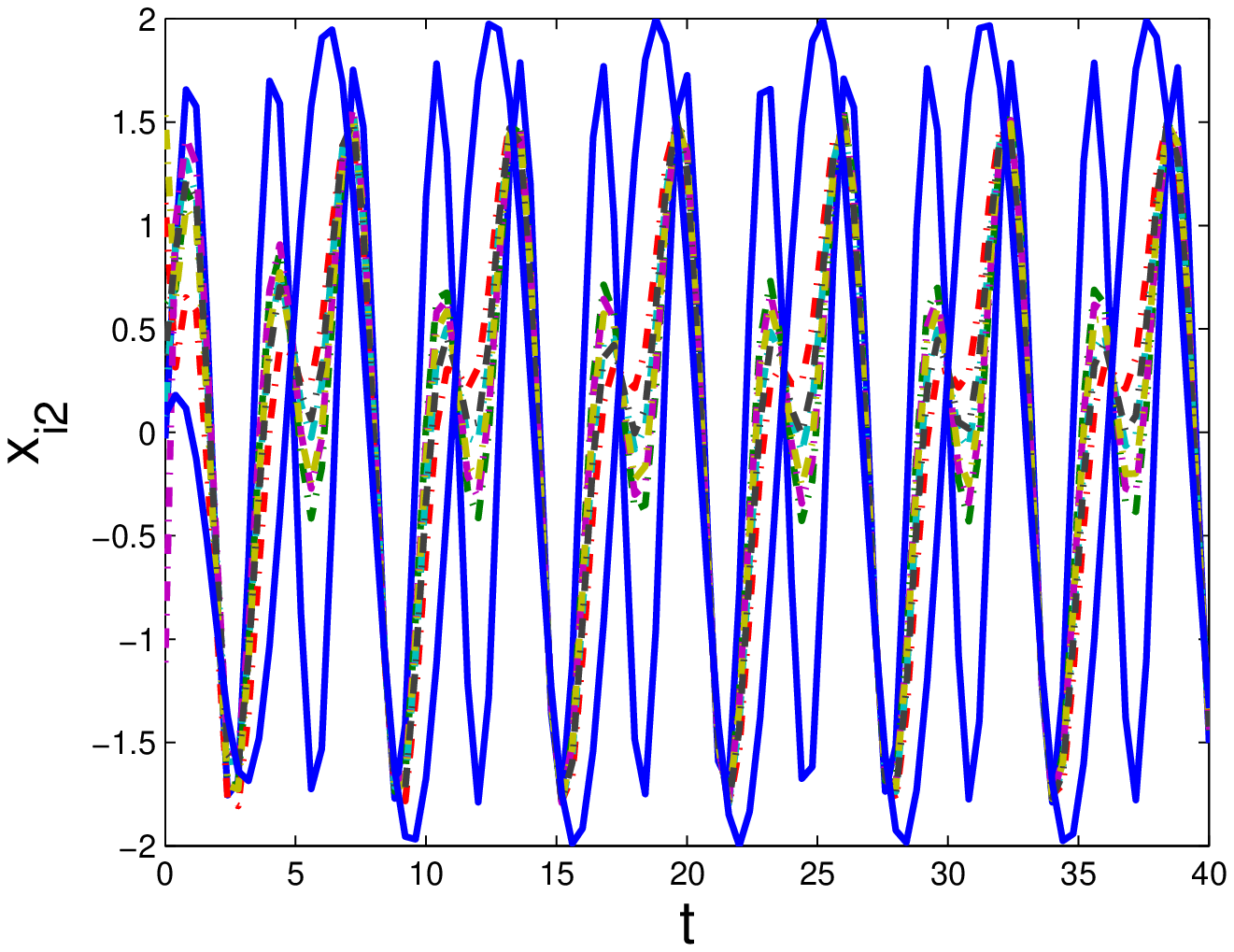}
\caption{The state trajectories of the agents. The solid and
dashdotted lines denote, respectively, the trajectories of the leaders and the
followers.}
\end{figure}

Solving the LMI \dref{alg1} by using the Sedumi toolbox \cite{sturm1999using}
gives the gain matrices $K$ and $\Gamma$ in \dref{acons} as
$$K=-\begin{bmatrix} 1.6203 & 4.7567\end{bmatrix},\quad
\Gamma=\begin{bmatrix} 2.6255  &  7.7075\\
    7.7075 &  22.6266\end{bmatrix}.$$
To illustrate Theorem 3,
select $\kappa=0.1$, $\varphi_i=0.005$,
and $\tau_i=5$, $i=2,\cdots,7$, in \dref{acons}.
The state trajectories $x_i(t)$ of the agents under
\dref{acons} designed as above are depicted in Figure 2,
implying that the containment control problem
is indeed solved.
The coupling gains
$d_{i}$ associated with the followers are drawn in Figure 3, which
are clearly bounded.

\begin{figure}[htbp]\centering
\centering
\includegraphics[height=0.3\linewidth,width=0.45\linewidth]{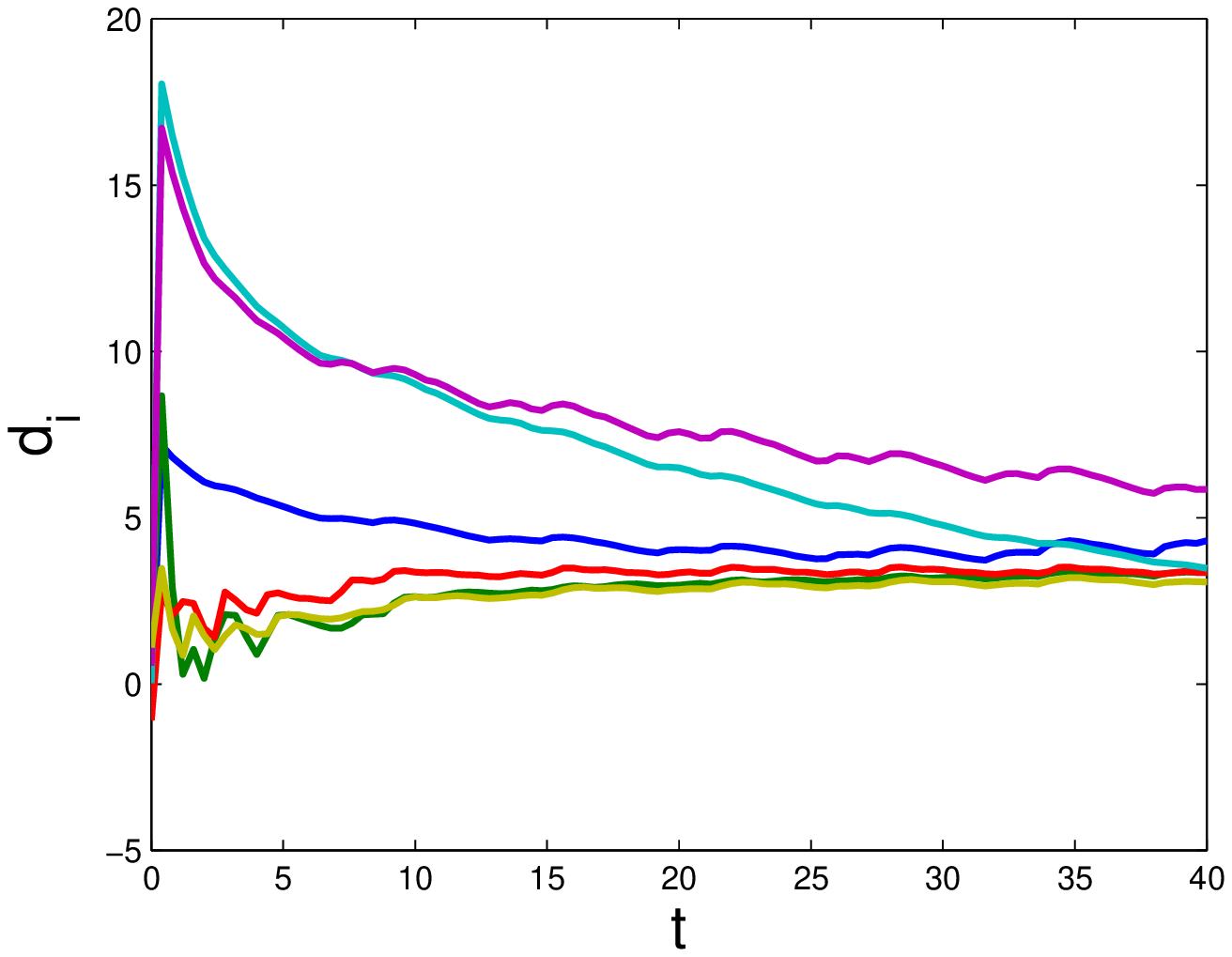}
\caption{The coupling gains $d_{i}$ in \dref{acons}.}
\end{figure}

\section{Conclusion}

In this paper, we have considered the
containment control problem for
multi-agent systems with general linear dynamics and
multiple leaders whose control inputs
are possibly nonzero and time varying. Based on the relative states
and relative estimates of the states of neighboring agents,
distributed static and adaptive continuous
controllers have been designed,
under which the containment error is uniformly ultimately
bounded, if
the subgraph associated with
the followers is undirected and for each
follower there exists at least one leader that has a directed path
to that follower.
A sufficient condition for the existence of these
containment controllers is that each agent is stabilizable and
detectable.
An interesting future topic is to consider the distributed
containment problem for the case with
general directed communication graphs.

\section*{Acknowledgements}
This work was supported by the National Natural Science Foundation
of China under grants 61104153 and 61225013,
National Science Foundation under
CAREER Award ECCS-1213291.


{\small
\begin{thebibliography}{10}

\bibitem{olfati-saber2004consensus}
R.~Olfati-Saber and R.~Murray, ``{Consensus problems in networks of agents with
  switching topology and time-delays},'' {\em IEEE Transactions on Automatic
  Control}, vol.~49, no.~9, pp.~1520--1533, 2004.

\bibitem{ren2007information}
W.~Ren, R.~Beard, and E.~Atkins, ``{Information consensus in multivehicle
  cooperative control},'' {\em IEEE Control Systems Magazine}, vol.~27, no.~2,
  pp.~71--82, 2007.

\bibitem{abdessameud2010consensus}
A.~Abdessameud and A.~Tayebi, ``On consensus algorithms for double-integrator
  dynamics without velocity measurements and with input constraints,'' {\em
  Systems \& Control Letters}, vol.~59, no.~12, pp.~812--821, 2010.

\bibitem{li2010consensus}
Z.~Li, Z.~Duan, G.~Chen, and L.~Huang, ``Consensus of multiagent systems and
  synchronization of complex networks: A unified viewpoint,'' {\em IEEE
  Transactions on Circuits and Systems I: Regular Papers}, vol.~57, no.~1,
  pp.~213--224, 2010.

\bibitem{li2011dynamic}
Z.~Li, Z.~Duan, and G.~Chen, ``Dynamic consensus of linear multi-agent
  systems,'' {\em IET Control Theory and Applications}, vol.~5, no.~1,
  pp.~19--28, 2011.

\bibitem{zhang2011optimal}
H.~Zhang, F.~Lewis, and A.~Das, ``Optimal design for synchronization of
  cooperative systems: State feedback, observer, and output feedback,'' {\em
  IEEE Transactions on Automatic Control}, vol.~56, no.~8, pp.~1948--1952,
  2011.

\bibitem{zhang2011fast}
H.~Zhang, M.~Chen, and G.~Stan, ``Fast consensus via predictive pinning
  control,'' {\em IEEE Transactions on Circuits and Systems I: Regular Papers},
  vol.~58, no.~9, pp.~2247--2258, 2011.

\bibitem{you2011network}
K.~You and L.~Xie, ``Network topology and communication data rate for
  consensusability of discrete-time multi-agent systems,'' {\em IEEE
  Transactions on Automatic Control}, vol.~56, no.~10, pp.~2262--2275, 2011.

\bibitem{li2011trackingTAC}
Z.~Li, X.~Liu, W.~Ren, and L.~Xie, ``Distributed tracking control for linear
  multi-agent systems with a leader of bounded unknown input,'' {\em IEEE
  Transactions on Automatic Control}, vol.~58, no.~2, pp.~518--523, 2013.

\bibitem{li2012adaptiveauto}
Z.~Li, W.~Ren, X.~Liu, and L.~Xie, ``{Distributed consensus of linear
  multi-agent systems with adaptive dynamic protocols},'' {\em Automatica}, in
  press, 2013.

\bibitem{grip2012output}
H.~Grip, T.~Yang, A.~Saberi, and A.~A. Stoorvogel, ``Output synchronization for
  heterogeneous networks of non-introspective agents,'' {\em Automatica},
  vol.~48, no.~10, pp.~2444--2453, 2012.

\bibitem{ji2008containment}
M.~Ji, G.~Ferrari-Trecate, M.~Egerstedt, and A.~Buffa, ``{Containment control
  in mobile networks},'' {\em IEEE Transactions on Automatic Control}, vol.~53,
  no.~8, pp.~1972--1975, 2008.

\bibitem{cao2011distributed2}
Y.~Cao, D.~Stuart, W.~Ren, and Z.~Meng, ``Distributed containment control for
  multiple autonomous vehicles with double-integrator dynamics: Algorithms and
  experiments,'' {\em IEEE Transactions on Control Systems Technology},
  vol.~19, no.~4, pp.~929--938, 2011.

\bibitem{cao2009containment}
Y.~Cao and W.~Ren, ``{Containment control with multiple stationary or dynamic
  leaders under a directed interaction graph},'' in {\em Proceedings of the
  48th IEEE Conference on Decision and Control and the 28th Chinese Control
  Conference}, pp.~3014--3019, 2009.

\bibitem{cao2012distributed}
Y.~Cao, W.~Ren, and M.~Egerstedt, ``Distributed containment control with
  multiple stationary or dynamic leaders in fixed and switching directed
  networks,'' {\em Automatica}, vol.~48, no.~8, pp.~1586--1597, 2012.

\bibitem{lou2012target}
Y.~Lou and Y.~Hong, ``Target containment control of multi-agent systems with
  random switching interconnection topologies,'' {\em Automatica}, vol.~48,
  no.~5, pp.~879--885, 2012.

\bibitem{galbusera2013hybrid}
L.~Galbusera, G.~Ferrari-Trecate, and R.~Scattolini, ``A hybrid model
  predictive control scheme for containment and distributed sensing in
  multi-agent systems,'' {\em Systems \& Control Letters}, vol.~62, no.~5,
  pp.~413--419, 2013.

\bibitem{mei2012distributed}
J.~Mei, W.~Ren, and G.~Ma, ``Distributed containment control for lagrangian
  networks with parametric uncertainties under a directed graph,'' {\em
  Automatica}, vol.~48, no.~4, pp.~653--659, 2012.

\bibitem{dimarogonas2009leader}
D.~Dimarogonas, P.~Tsiotras, and K.~Kyriakopoulos, ``{Leader--follower
  cooperative attitude control of multiple rigid bodies},'' {\em Systems and
  Control Letters}, vol.~58, no.~6, pp.~429--435, 2009.

\bibitem{meng2010distributed}
Z.~Meng, W.~Ren, and Z.~You, ``{Distributed finite-time attitude containment
  control for multiple rigid bodies},'' {\em Automatica}, vol.~46, no.~12,
  pp.~2092--2099, 2010.

\bibitem{li2011containment}
Z.~Li, W.~Ren, X.~Liu, and M.~Fu, ``{Distributed containment control of
  multi-agent systems with general linear dynamics in the presence of multiple
  leaders},'' {\em International Journal of Robust and Control}, vol.~23,
  no.~5, pp.~534--547, 2013.

\bibitem{agaev2005spectra}
R.~Agaev and P.~Chebotarev, ``On the spectra of nonsymmetric laplacian
  matrices,'' {\em Linear Algebra and its Applications}, vol.~399, no.~1,
  pp.~157--178, 2005.

\bibitem{ren2005consensus}
W.~Ren and R.~Beard, ``{Consensus seeking in multiagent systems under
  dynamically changing interaction topologies},'' {\em IEEE Transactions on
  Automatic Control}, vol.~50, no.~5, pp.~655--661, 2005.

\bibitem{shevitz1994lyapunov}
D.~Shevitz and B.~Paden, ``Lyapunov stability theory of nonsmooth systems,''
  {\em IEEE Transactions on Automatic Control}, vol.~39, no.~9, pp.~1910--1914,
  1994.

\bibitem{young1999control}
K.~Young, V.~Utkin, and U.~Ozguner, ``A control engineer's guide to sliding
  mode control,'' {\em IEEE Transactions on Control Systems Technology},
  vol.~7, no.~3, pp.~328--342, 1999.

\bibitem{edwards1998sliding}
C.~Edwards and S.~Spurgeon, {\em Sliding Mode Control: Theory and
  Applications}.
\newblock London: Taylor \& Francis, 1998.

\bibitem{yang2012semi}
T.~Yang, A.~A. Stoorvogel, H.~Grip, and A.~Saberi, ``Semi-global regulation of
  output synchronization for heterogeneous networks of non-introspective,
  invertible agents subject to actuator saturation,'' {\em International
  Journal of Robust and Nonlinear Control}, in press, 2012.

\bibitem{meng2013global}
Z.~Meng, Z.~Zhao, and Z.~Lin, ``On global leader-following consensus of
  identical linear dynamic systems subject to actuator saturation,'' {\em
  Systems \& Control Letters}, vol.~62, no.~2, pp.~132--142, 2013.

\bibitem{khalil2002nonlinear}
H.~Khalil, {\em Nonlinear Systems}.
\newblock Englewood Cliffs, NJ: Prentice Hall, 2002.

\bibitem{corless1981continuous}
M.~Corless and G.~Leitmann, ``Continuous state feedback guaranteeing uniform
  ultimate boundedness for uncertain dynamic systems,'' {\em IEEE Transactions
  on Automatic Control}, vol.~26, no.~5, pp.~1139--1144, 1981.

\bibitem{ioannou1984instability}
P.~Ioannou and P.~Kokotovic, ``Instability analysis and improvement of
  robustness of adaptive control,'' {\em Automatica}, vol.~20, no.~5,
  pp.~583--594, 1984.

\bibitem{boyd1994linear}
S.~Boyd, L.~El~Ghaoui, E.~Feron, and V.~Balakrishnan, {\em {Linear Matrix
  Inequalities in System and Control Theory}}.
\newblock Philadelphia, PA: SIAM, 1994.


\bibitem{sturm1999using}
J.~Sturm, ``{Using SeDuMi 1.02, a MATLAB toolbox for optimization over
  symmetric cones},'' {\em Optimization Methods and Software}, vol.~11, no.~1,
  pp.~625--653, 1999.

\end{thebibliography}
}
\end{document}